\documentstyle[12pt,aaspp4]{article}			

\newcommand{\GHz}{$\,{\rm GHz}\,$}
\newcommand{\registered}
   {{\ooalign{\hfil\raise0.07ex\hbox{\sc r}\hfil%
              \crcr\mathhexbox20D}}}

\begin{document}

\title{An Absolute Flux Density Measurement of the Supernova Remnant \\ Cassiopeia A at 32 GHz}
\author{Brian S. Mason \altaffilmark{1,2} , Erik M. Leitch \altaffilmark{3,4} , Steven T. Myers \altaffilmark{1,5} ,  
 John K. Cartwright\altaffilmark{3}, and A.C.S. Readhead \altaffilmark{3}}

\altaffiltext{1}{\footnotesize University of Pennsylvania, 209 S. 33rd St., Philadelphia, PA 19104-6396}
\altaffiltext{2}{\footnotesize Current Address: California Institute of Technology, 105-24, Pasadena, CA 91125}
\altaffiltext{3}{\footnotesize California Institute of Technology, 105-24,  Pasadena, CA 91125 }
\altaffiltext{4}{\footnotesize Current Address:  University of Chicago, 5640 S. Ellis Ave., Chicago, IL 60637}
\altaffiltext{5}{\footnotesize Current Address: NRAO, Socorro, NM 87801}

\begin{abstract}

We report 32 GHz absolute flux density measurements of the supernova remnant 
Cas A, with an accuracy of 2.5\%.  The measurements were made with the 1.5-meter telescope
at the Owens Valley Radio Observatory.  The antenna gain had been measured by NIST
in May 1990 to be $0.505 \pm 0.007 \frac{{\rm mK}}{{\rm Jy}}$.  Our observations of Cas A
in May 1998 yield $S_{cas,1998} = 194 \pm 5 \,\,{\rm Jy}$.  We also report
absolute flux density measurements of 3C48, 3C147, 3C286, Jupiter, Saturn
and Mars.

\end{abstract}

\keywords{ standards --- ISM: individual (Cassiopeia A) --- 
planets and satellites: individual (Jupiter, Mars, Saturn) ---
quasars: individual (3C48, 3C147, 3C273, 3C286, 3C345) ---
galaxies: individual (3C84, 3C218, 3C353) }

\section{Introduction}
\label{sec:intro}
Absolute flux density calibration is an important issue for all astronomical observations.
Most absolute calibration efforts have occurred at lower 
frequencies ($\lesssim 10\,$ GHz) since it is easier to measure the far-field properties
of antennas in this regime, or to use feed systems of calculable gain.
See, for example,  Conway et al. (1963)  , Kellermann (1964), Kellermann et al.  (1969),
 Baars (1977)
and references therein.   Most of these flux density scales are based upon measurements of 
the supernova remnant Cassiopeia A (Cas A) with horn-type or dipole antennas.
  At higher frequencies this endeavor is substantially more difficult, and investigators
have often used flux densities extrapolated from low frequencies or computed from
theoretical models of celestial sources.  One exception is the work of Wrixon~(1972),
who measured the brightness temperature of Jupiter at frequencies between 20.5 and 33.5 GHz
with a 20 foot antenna
at the Hat Creek observatory; this telescope was calibrated with a
 far-field transmission/reception
measurement.  This measurement was the basis of the calibration used for the SZE
measurements of Myers et al. (1998), but given the relatively large ($5.5\%\,$) uncertainty
in Wrixon's $T_{J}$ it is desirable to improve upon this.
This paper presents an {\it absolute} measurement with an accuracy of $2.5\%\,$
of the flux density of Cas A at 32 GHz, a 
 calibrator commonly used at all frequencies and the second brightest extrasolar source in the sky
in our observing band.  

While absolute calibration is an important consideration for many different types of 
experiments, our calibration efforts
were undertaken with the particular intent of supporting the intrinsic CMB and Sunyaev-Zeldovich
programs at the Owens Valley Radio Observatory (OVRO) and other similar observations. 
 Absolute
calibration is an especial concern for Sunyaev-Zeldovich studies since the error in the quantity of 
interest ($H_o$) is proportional to twice the calibration error.  
It is of great importance that the calibrations of intrinsic CMB anisotropy experiments be as accurate as
possible in order to place more stringent limits on the parameters of cosmology.
Several
published CMB experiments use Cas A as their primary calibrator,
such as the Saskatoon experiment (Netterfield et al., 1996) and
CAT (Scott et al., 1996).
These experiments typically have calibration uncertainties 
$\gtrsim \, 10\%$.
It is important to reduce these uncertainties as much as possible and
to use a single flux density scale at or near 32 GHz.  This affects a number
of experiments including the RING5M (Leitch et al., 1999), MAT (Torbet et al. 1999; Miller et al.
1999), 
QMAP (Devlin et al., 1998),
 CBI (Readhead et al., 1998) , DASI (White et al., 1998) and the VSA (Jones and Scott,
1998).

The 1.5-m telescope used in the present observations
was originally designed and constructed by JPL as part of 
the Deep Space Network (DSN) program.
The telescope has a parabolic primary reflector with an off-axis
secondary and a dual feed horn, and a Dicke-switching receiver at the Cassegrain focus.
To establish absolute gain standards for DSN,
the National Institute of Standards and Technology (NIST) was employed to measure the 
aperture efficiency of the 1.5-m antenna; these measurements are discussed
in \S~\ref{sec:nist}.  OVRO obtained the instrument
in 1993, and our current calibration program commenced in 1996, with efforts
focused on characterizing the receiver and reworking the telescope control
code.  
Most of the useful astronomical observations were taken during the spring
of 1998 (from early March through the end of May).

We selected Cassiopeia A  ($\alpha\,=\,23^{\rm h}23^{\rm m} 26^{\rm s}.920$  ,  $\delta\,=\,+58^{\circ}49'07''.50$, J2000)
as our primary calibration source because of its brightness and accessibility
from the latitude of the observatory. 
Observations of other calibrator sources relative to Cas A were carried out in
the fall of 1998, as well as during epochs previous to our absolute measurements, allowing
the determination of absolute flux densities for these sources as well.

In the following, we first present a brief description of the
NIST calibration measurements (\S~\ref{sec:nist}) and our own 
characterization of the instrument (\S~\ref{sec:rxcal});
 following this is a summary of observations
taken at OVRO in the spring of 1998 (\S~\ref{sec:obs}), and a detailed discussion of our error
budget (\S~\ref{sec:syserr}).  Next we describe observations of 3C48, 3C147, 3C286, Jupiter, Mars, Saturn and other sources on the OVRO 5.5-meter telescope relative to Cas A, employing our absolute value for $S_{\it cas}$
($\S~\ref{sec:extendedscale}$).
Finally we summarize our
findings and compare them to the results of other flux density scales
(\S~\ref{sec:concl}).

\section{Antenna Characterization: NIST Measurements}
\label{sec:nist}

Flux density, $S_{\nu}$, is related to the spectral power, $P_{\nu}$, by
\begin{equation}
\label{eq:power}
P_\nu = A_{eff}(\nu)  \times S_\nu ,
\end{equation}
where
 $A_{\it eff}(\nu)$ is the {\it effective area} of the telescope.  $A_{\it eff}$
is always less than the geometrical area of the telescope for a well-designed
system due to diffraction at the edge of the dish, 
surface irregularities, scattering off
of the secondary support structure (not a problem for unblocked apertures
such as this one), slight impedance mismatches, and so forth.
The most difficult aspect of most absolute flux density 
measurements at high frequency is the 
accurate assessment of $A_{\it eff}$.

It can be shown from elementary thermodynamic considerations (e.g., Rohlfs and Wilson, 1996;
Krauss, 1986) that 
\begin{equation}
\label{eq:radiation}
\Omega_{Ant} \times A_{eff} = \lambda^2
\end{equation}
where $\lambda$ is the wavelength of the incident radiation and
\begin{equation}
\label{eq:omegaant}
\Omega_{Ant} = \int\!\!\!\int d\phi \, d(\cos{\theta}) P_{Ant}(\theta,\phi).
\end{equation}
Here $P_{\it Ant}(\theta,\phi)$ is the response function of the antenna normalized to one at
the maximum.  If we define the {\it gain}, $g\,$, of a telescope in a given direction
 as the ratio of $P_{\it Ant}$ in that direction to the mean value of $P_{\it Ant}$
on the sphere, then it can be shown that 
\begin{equation}
\label{eq:gain}
g|_{max} = \frac{4 \pi}{\Omega_{Ant}}.
\end{equation}
Since $g$ is a ratio, it is often expressed in decibels (dBi, or decibels above the isotropic
level implied by the spherical average).

It is convenient to express spectral power in terms of the {\it antenna temperature} $T_{\it Ant}$,
defined as the temperature a beam-filling blackbody would need to have (in the
Rayleigh-Jeans limit) in order to produce the observed spectral power.  The antenna
system may then be characterized by the Kelvin per Jansky {\it sensitivity}, $\Gamma$:
\begin{equation}
\label{eq:sensitivity}
\Gamma = \frac{\lambda^2}{8 \pi k} \times g|_{max}.
\end{equation}
Numerically, this is 
\begin{equation} 
\label{sensitivity2}
\Gamma = 0.362 \times \frac{A_{eff}}{{\rm meters^{2}}} \frac{{\rm mK}}{{\rm Jy}}.
\end{equation}
In the far-field of a given antenna, these properties ($\Gamma$ , $A_{\it  eff}$, 
$\Omega_{\it Ant}$) are independent of distance.  The far-field is conventionally
defined to be the distance at which the phase variation (across the aperture
of the antenna) of an electric field originating
at a point at distance $r_{\it ff}$ is less than 22.5$^{\circ}$; this criterion
yields a beam pattern which is, to a good approximation, independent of the
distance of the illuminating source from the antenna.
For a flat aperture the given phase constraint results in
 $r_{\it ff} > 2 D^2 / \lambda$. 

Since for many telescopes the far-field is inconveniently far
away, the direct measurement of far-field antenna characteristics often requires
measurements to be made over long distances outside of a controlled laboratory setting.
In this context, boresighting, ground reflections and atmospheric attenuation are
all difficult issues to deal with.  
Extensive work has consequently been done at NIST and elsewhere to
 develop near-field techniques 
to measure the far-field power pattern and gain of antennas;  this work and the theory
behind it is summarized in Baird et al. (1988) Lo and Lee (1993) Newell et al. (1988) Rahmat-Samii (1993).
  The technique employs
a measurement of the near-field electric field of the antenna; the resulting field distribution
(after correcting for the probe antenna characteristics) is related to 
the far-field electric field by a Fourier transform.  Ranged measurements are used to
fit out mutual coupling terms due to imperfect impedance matching between the target
and probe antennas (Lo and Lee, 1993 ; Francis, 1990); the probe gains themselves
are determined using a three-antenna technique (Lo and Lee, 1993).
 In May of 1990, our telescope was characterized at NIST's near field test range in 
Boulder, Colorado  (NIST, 1990). 
  The compactness of the antenna permitted this procedure
to be carried out without disassembling the instrument for shipping, greatly
reducing uncertainties which may otherwise arise due to the relative displacement of the various
optical components of the system; for more discussion of this point,
see \S~\ref{sec:syserr}.  NIST assigns a $1.4 \% \, $ ($1-\sigma$) systematic error to
this measurement in the calibration report; this uncertainty is dominated by the determination
of the probe antenna gain and normalization amplitude.   As discussed in \S~\ref{sec:rxcal},
the 1.5-meter is outfitted with a Dicke-switching receiver with two horns.
The gains of these horns were
separately measured; we designate them  the ANT and REF horns.  The NIST gains, however, are referred to the {\it back} of the feed horns.  This implies that the stated gains include the effects of ohmic losses in the feeds, whereas our calibration procedure effectively removes these losses.  We will account for this in our final systematic error budet.

\begin{table}[t!]
\begin{center}
\begin{tabular}{ l l l l l r } 
\tableline\tableline
$\nu$ (\GHz) & g ($\, {\rm dBi} \,$)  & g ($10^5$) & $A_{\it eff}$ (${\rm m}^2$)  &  $\eta_A$ & G (mK/Jy)\\  
\tableline
31.8 (ANT) & $52.54 \pm 0.06$ & 1.795 & 1.270 &  71.9\% & $0.460 \pm 0.006$ \\  
32.0 (ANT)  & $53.04 \pm 0.06$ & 2.014 & 1.407 & 79.6\% & $0.509 \pm 0.007$ \\  
32.0 (REF) & $52.96 \pm 0.06$ &  1.977 & 1.381 & 78.1\% & $0.500 \pm 0.007$ \\  
32.3  (ANT) & $53.32 \pm 0.06$ & 2.148 & 1.472 &  83.3\% & $0.533 \pm 0.007$ \\   \tableline
\end{tabular}
\end{center}
\caption{NIST antenna gain measurements}
\label{tbl:nistgain}
\end{table}

The ANT beam gain was measured at 31.8, 32.0 and 32.3 GHz ; the REF beam gain
was only measured at 32.0 GHz.  These measurements are summarized in 
Table~\ref{tbl:nistgain}.  Note that the measured gains at the band edges differ by
$13.6\%$; while the nominal receiver bandpass is 3 GHz, it is clear that the extrapolation
of the observed gains to 30.5 and 33.5 GHz will introduce a very large uncertainty.  Furthermore,
the REF beam was only characterized at the band center.  For these reasons an additional
filter was introduced into the signal path, reducing our system bandpass to 500 MHz centered
on 32.0 GHz (see \S~\ref{sec:rxcal}).  We therefore adopt the mean of the ANT and REF
beam gains at 32.0 GHz, $g = (1.995 \pm 0.028) \times 10^{5}$ ($53.00 \pm 0.06\,$ dBi),
 corresponding to
 $\Omega_{\it Ant} = (6.298 \pm 0.088) \times 10^{-5}\,$~Sr,
$A_{\it eff} = (1.394 \pm 0.020) \, {\rm meters^{2}}$ and
 $\Gamma =(0.505 \pm 0.007) \frac{{\rm mK}}{{\rm Jy}}$.  Assuming a geometrical
area of $1.77 \, {\rm meters^2}\,$, this yields an aperture efficiency of
 $(78.8 \pm 1.1) \%$. 
The average of the ANT and
REF gains is the appropriate quantity to use in the context of our double-differenced FLUX
procedure (see \S~\ref{sec:obs}).

NIST also measured the beam patterns of both ANT and REF beams; the
beams were found to be well-described by Gaussians of 
$25'.8 \pm 0'.6$  FWHM, corresponding to $\sigma =
10'.96 \pm 0'.25$.  
Integrating a Gaussian of this width out to the radius of the first
measured nulls in the beam pattern
(at $\pm 24'.0$), we compute
$\Omega_{\it Beam} = (5.80 \pm 0.26) \times 10^{-5}$~Sr and a beam
efficiency of
\begin{eqnarray*}
\label{eq:beamefficiency}
 \eta_{Beam} &=& \frac{\Omega_{main}}{\Omega_{Ant}} \\
	     &=& 87.0 \pm 4.8 \% .
\end{eqnarray*}
The beam throw was measured to be $1^{\circ}.00 \pm 0^{\circ}.01$ , with
a $0^{\circ}.02 \pm 0^{\circ}.1$ ($=EL_{\it ANT} - EL_{\it REF}$) elevation offset.
The relatively high beam and aperture efficiencies demonstrate
the efficacy of the clear-aperture design and small, stiff primary dish.

It is likely that the 1.8\% difference
between the ANT and the REF gains is due to a slight impedance mismatch
on the REF side of the signal chain;  NIST measured the return loss on the
REF horn to be several dB lower than that on the ANT horn.  This could be due
to differential isolation in the Dicke switch.

\section{Receiver Characterization and Antenna Temperature Calibration}
\label{sec:rxcal}

The receiver is a standard Dicke-switching two-horn receiver.
Celestial signals enter the receiver via these two scalar feeds (uncooled)
and pass into the 15 K dewar where right circular polarization is 
selected.
A cross-guide coupler before the Dicke-switch permits the injection
of a calibration (cal) signal.  The Dicke-switch itself operated
at 1 kHz during observations, providing alternating 0.5 msec ``integrations''
against each feed.
The switched signal is amplified by a HEMT (+28 dB of gain), 
the last component in the 15 K dewar.  The 15 K cryogenic stage
 displayed remarkable stability from fall 1996 through spring 1998.

The 15 K dewar is enclosed in a dewar which is maintained at $\sim 70$ K.
On entering this stage, the RF passes first through an Avantek amplifier
(second stage, +22 dB) and then through a double-sideband mixer, which 
converts the signal down to a 0 to 1.5 GHz IF.
Before leaving the dewar, the
IF is amplified and passed through a 1.5 GHz low-pass filter.
This signal is piped via coaxial cable to a climate-controlled trailer where
the detection and post-detection signal processing is carried out.
For the reasons detailed in \S~\ref{sec:nist}, detection of the full band power
is undesirable, so this signal was passed through a 250 MHz low-pass filter,
giving a system bandpass of 500 MHz on the sky;
 the resulting signal is split, half going
to a diode detector, which feeds a Lock-in-Amplifier (LIA) for the detection of switched power,
 and half going to an HP 437B power meter for detection of total power.  In order to
boost the IF signal well above the lowest levels that the LIA was capable of detecting,
another IF amplifier (+17 dB of gain) was inserted in the signal chain after the coaxial
cable and before the final low-pass filter.

The linearities of the power meter power head and diode detector were assessed by measuring
the cal signal increment with an ambient-temperature microwave absorber
 placed in
front of the feed-horns and varying the attenuation prior to the powerhead and detector.
The diode detector's departures
from linearity over the dynamic range encountered in calibration observations is less
than $0.1 \%$; the power head's deviation is less than $0.2\%$.   These elements are
thus not significant sources of nonlinearity.  The receiver on the other hand is somewhat
nonlinear, which is discussed below.

The Dicke switch isolation was measured by locking the switch onto the ANT feed while
repeatedly firing the cal signal into the REF feed.  In 112 pairs of on/off
averages of the system power, a mean fractional increase 
in the system temperature of $(5.7 \pm 1.3) \times 10^{-4}$ was
observed.  These tests were conducted against an external cold
load for stability; assuming $T_{\it RX} = 55 \, {\rm K} + 77 \,  {\rm K} = 132 \, {\rm K}\,$ 
and $T_{\it cal} = 14.34 \,{\rm K}$ (see below) , this implies a leakage of $0.54 \pm 0.11 \%$
of the cal signal.  This is equivalent to an isolation of $-22.7 \pm 0.9$ 
dB.  Any effects such as losses or a finite duty cycle in the Dicke switch 
do not affect our results as they will affect celestial observations
and observations of our calibration loads in the same fashion (see below).

Although the receiver design incorporates several mechanical waveguide switches which permit
some degree of control over the magnitude of the cal diode signal, these switches
were not exercised {\it at any point} during or between our calibration measurements
and observations.  This precaution was deemed necessary due to the possibility that the
switches might not toggle between two precisely defined impedance states,
which would cause the apparent cal signal brightness to change.
The receiver does not have any variable attenuators, which would have to be
calibrated at a number of settings.

Although the NIST calibration allows the accurate computation of flux densities given $T_{\it Ant}$,
the task of determining $T_{\it Ant}$ from observed quantities (milliVolts) is not trivial.  This
issue is more important in the context of an absolute measurement, since for experiments
which use an external flux density reference the sensitivity (K/Jy) and the receiver gain (mV/K) are
 not determined independently:   the overall gain (mV/Jy, or some equivalent quantity)
is obtained directly.   Consequently, great care was taken in designing
a power ($T_{\it Ant}$) calibration scheme.  This task is made somewhat easier
by the fact that we measure the antenna temperature in units of the cal signal, whose
derived magnitude is not highly sensitive to slight errors in our brightness temperature
references (two microwave absorbers at known
thermodynamic temperatures-- see \S~\ref{sec:syserr}, and below).

To measure the receiver temperature $T_{\it RX}$ of our system and the effective
brightness temperature $T_{\it cal}$ of our cal signal, we use the standard
hot-cold load scheme.  Our hot load is an ambient temperature microwave 
absorber and our cold load is an absorber immersed in liquid nitrogen.
The increment in total power is measured against each of these loads
 providing a measurement
of $T_{\it cal}$, $T_{\it RX}$ and   the receiver nonlinearity.
The loads themselves are our ultimate temperature
references.   In terms of the temperatures of the hot and cold loads, the
receiver temperature $T_{\it RX}$ 
is given by
\begin{equation}
T_{\it RX} = \frac{T_{hot} - y\,T_{cold} }{y-1}
\end{equation}
where $y$ is the observed ratio of total power against the hot load
to total power against the cold load.  Once $T_{\it RX}$ is known,
the effective brightness temperature of the cal diode may be determined
by observing the total power against either of  the hot or cold loads with
the cal diode on and off.  For observations against the cold load, we find
\begin{equation}
\label{eq:tcalcold}
T_{\it cal} = \frac{y_{\it cal} - 1}{y-1} \, (T_{hot} - T_{cold})
\end{equation}
where $y_{\it cal}$ is the ratio of the total power observed with the
cal diode on to total power observed with the cal diode off.  If $T_{\it cal}$ is
derived from observations against the hot load, we find
\begin{equation}
T_{\it cal} = \frac{y_{\it cal} - 1}{y-1} \, y\,(T_{hot} - T_{cold}).
\end{equation}
Since observations at lower power levels are
less affected by increased system noise and nonlinearity (see below), we derive
$T_{\it cal}$ from observations against the cold load.

The microwave absorber employed
for each load was a 3-inch thick sheet of 
ECCOSORB CV\footnote{ECCOSORB is a registered 
trademark of Emerson and Cuming Microwave Products.}.
Tests at liquid nitrogen
temperature indicate that the absorber had an optical depth $ \tau > 7.4$.
At ambient temperatures, the optical depth is expected to be higher
(Hemmati, 1985), but this measurement is more difficult due to the increased
system noise.
The cold load was assumed to have a thermodynamic temperature equal to
the boiling point of liquid nitrogen (76.3 K at atmospheric pressures characteristic
of the Owens Valley).  The front-face temperature of the hot load was measured
with a hand-held infrared thermometer; this thermometer was calibrated (using
ECCOSORB-CV) against
a platinum resistance thermometer with a specified accuracy at room temperature
of $\pm 0.02\, {\rm K}\,$.  While in principle this calibration may have been necessary
to relate the thermodynamic temperature to the effective infrared brightness temperature
over the IR thermometer's bandpass (i.e., to correct for the unknown and presumably finite
IR emissivity of ECCOSORB-CV) we found that both thermometers measured the same temperatures
to $\sim 0.1 {\rm K}\,$.
Assuming that $\tau \ge  7.4\,$ for our absorber, the
thermodynamic and brightness temperatures for both loads are identical to better than $0.1\%$.
Given our measured optical depth, temperature gradients across the hot load do not significantly
affect our measurements provided we measure the temperature of the front face of the
absorber.

The
manufacturer's specifications indicate that reflections from the absorber 
at 32 GHz are suppressed
by 50 dB.  The cold load cooler is constructed of a cross-linked polyethylene foam,
which possesses good thermal properties and has a very low optical depth
at microwave frequencies ($\tau \sim 7 \times 10^{-3}$).  Laboratory measurements
of the cooler show that it evinces a brightness temperature $\sim 0.7\,$ K higher
than a cone of absorber dipped in liquid nitrogen.  This measured brightness temperature excess is also
consistent with the expected brightness temperature of a 300 K emitter with an optical
depth of $\tau = 7 \times 10^{-3}$. We correct the cold load brightness
temperature for this emission.
In spite of the simplicity of the cone-load arrangement,
these loads were
not used for daily $T_{\it RX}$
and $T_{\it cal}$ measurements due to their lack of long-term emission stability.
One potential concern which arises in the usage of a liquid-nitrogen-filled cooler
is the impedance mismatch at the box-nitrogen interface; if there is significant power
being emitted out of the horns (presumably of order $T_{\it RX}$), then some fraction
of this power would be reflected back into the system, artificially increasing the
measured receiver temperature.
To the end of calibrating the first OVRO SZE experiment, Herbig et al. (1994)
constructed a cold load box of the same cross-linked polyethelyene foam with an impedance-matching
network on one side.  This network consisted of a hexagonal pattern of (liquid-nitrogen-filled) holes
on the inside of the cooler wall which provides a refractive index intermediate to that between
free space and liquid nitrogen.  No significant difference in power was observed between the
matched and unmatched sides of the cold load.  This together with the good agreement between
the measured and theoretically computed box emission, show that neither internal reflections
nor other excess emission mechanisms contribute significantly to our cold load measurement.
For a more detailed discussion on the effects
of excess emission on our $T_{ant}$ calibration, see \S~\ref{sec:syserr}.

In the taking of $T_{\it RX}$ measurements, a $\sim 5\%\,$ receiver nonlinearity is seen
(a compression of the cal signal at the hot load total power levels relative to what
is seen at cold load power levels).  Most likely this is due to a slight compression of
the second or third stage amplifiers. We correct our measurements for nonlinearity
by assuming that our system response $P_{obs}$ to an input power $P_{true}$
is described by
\begin{equation}
\label{eq:tpnl}
P_{obs} = g \times (1 + b P_{obs}) \times P_{true},
\end{equation}
where $g$ is the receiver gain and $b$ characterizes the nonlinearity in the system.  For small excursions
from perfect linearity as observed this should be a good description. 
 This is the same description employed by Leitch et al. (1999)
with good results on the OVRO 5.5-m Ka-band system, as well as the 40-m Ku-band system.

We measure $b$ on an observation-by-observation basis by equating the cal signal increments
observed against the hot and cold loads and solving for $b$.
It is not significant to quote a mean value for $b$ since this depends on $g$ 
at the time of the observation which is
significantly time variable.  Application of this correction to each data point reduced the mean
$T_{\it cal}$ by $2\%$, and $T_{\it RX}$ by $8\%$.

To measure the effectiveness of a nonlinearity correction based
on Eq.~\ref{eq:tpnl} , it is convenient to compare the cal diode
increments observed against hot and cold loads to the increment
observed against the sky at zenith.  For all the data taken in the 14 $T_{\it cal} \,$measurements,
prior to the application of the nonlinearity correction the ratios
of the raw cal increment observed against the hot
load to the raw cal increment observed at zenith, and the equivalent
quantity for the cold load, are $0.95 \pm 0.01$ and $0.99 \pm 0.01$
respectively.  After the application of a nonlinearity correction
derived separately for each observation, this ratio (for both loads)
is $ 1.00 \pm 0.02$.  Since no information about the cal signal increment
at zenith was employed in the derivation of $b$,
this is an independent indication that our simple model is a good 
description of the system non-linearity over the whole range of powers
relevant to our observations.

Using this arrangement we measured $T_{\it RX}$ and $T_{\it cal}$ on 14 separate occasions
during the period 10 May 1998 through 15 May 1998;  after correcting for receiver nonlinearity
and the Dicke-switch isolation, as well as applying a slight correction\footnote{This correction
amounted to a 0.4\% adjustment of our measured hot load temperatures, and a 0.8\% adjustment
of our assumed cold load temperature.}  for
deviations from the Rayleigh-Jeans law, we find $T_{\it RX} = 54.8 \pm 0.9 \,$ K and
 $T_{\it cal} = 14.34 \pm 0.07 \,$ K (measurement error only).  

\section{Observations}
\label{sec:obs}

The radiometry procedures used in observations on the 1.5-m telescope are the same
as those used on the other single-dish radio telescopes at OVRO.
The two fundamental procedures are FLUX and CAL measurements.  A FLUX
measurement is a double switched measurement in which the source is alternately
placed in the REF and ANT beams; two integrations of switched power are executed
in each position.  This cancels out constant and linear power backgrounds in the measurement,
such as those caused by the atmosphere, receiver noise and ground pickup.  A CAL
procedure consists of two integrations of switched power with the cal diode signal being injected
into one of the receiver arms, bracketed by two integrations of switched power with the 
cal signal off.  For more detail on these procedures refer to Readhead et al. (1989),
Myers et al. (1997), and Leitch et al. (1999).
It is additionally possible to record (singly) switched power; this procedure is known as an
AVERAGE.  By doing a set of AVERAGEs around the nominal pointing center of some source 
, it is possible to derive the pointing offsets; this is known as a POINT
procedure.  Generally four AVERAGEs were done spaced at $\pm 0^{\circ}.27\,$ in azimuth
and elevation around the source.

In March 1998 two days of pointing data were taken on bright celestial sources
(Cas A, the Crab and Venus).  These observations consisted of a cycle of POINT, FLUX,
and CAL procedures executed repeatedly against the source in long tracks.
 From these data a pointing model
was derived which reduced the rms pointing offset from 15'.4  to 2'.2  (0'.85
on Cas A) in addition to removing overall systematic pointing offsets.  This is the pointing model which
was used in subsequent observations.  Due to the weather conditions in May reliable pointing tests
were not possible, but for a small stiff dish such as this that presents no problem, as all of the
pointing errors are likely to be constant offsets.

On 22 April 1998, observations of Cas A consisting of alternate pointings on the ANT and
REF beam positions were taken in order to confirm the NIST-measured beamthrow. 
We found, relative to these fixed offsets, an azimuth separation of 
$0'.04 \pm 0'.12$ , and a zenith angle separation of $0'.70 \pm
0'.26$.
We take this to be consistent with the nominal $(\Delta {\rm AZ}, \Delta {\rm EL}) =
(-1^{\circ}.0 \pm 0^{\circ}.01,
0^{\circ}.0 \pm 0^{\circ}.1)$ offsets.  Given our 25'.8 FWHM main beam, 
the neglect of the 0'.7 ANT/REF zenith
angle offset 
amounts to a potential $0.3\%$ systematic error in our FLUX measurements.

During the period 10 May 1998 -- 16 May 1998, observations of Cas A were
conducted when the weather permitted.  These observations consisted
of FLUXes (10 second integrations in each segment, with 15 second idle
times to allow the telescope to acquire the position) interleaved
at 10 minute intervals with CAL's (10 second integrations, 5 second
idles).  The pointing model described above was employed.
Full tracks on the source were used in order to average out variations in atmospheric opacity or
residual ground spillover over the course of a track.

The data were calibrated by scaling the observed FLUXes by the ratio
of observed CAL's to the measured cal diode temperature
($T_{\it cal} = 14.34 \pm 0.07 \,{\rm K}$).  Since the CAL and FLUX measurements
were taken at the the same power level 
 no nonlinearity correction is necessary. 
If the standard deviation of any individual CAL or FLUX procedure is more than four times
the average standard deviation observed in the entire data set, that datum is rejected.
Similar $5-\sigma$ edits were performed on the CAL and FLUX means.  
These edits remove a negligible fraction of the data but
eliminate grossly discrepant data points caused by atmospheric interference 
 or rare instrumental glitches.

\begin{figure}
\vspace{3.2in}
\includegraphics{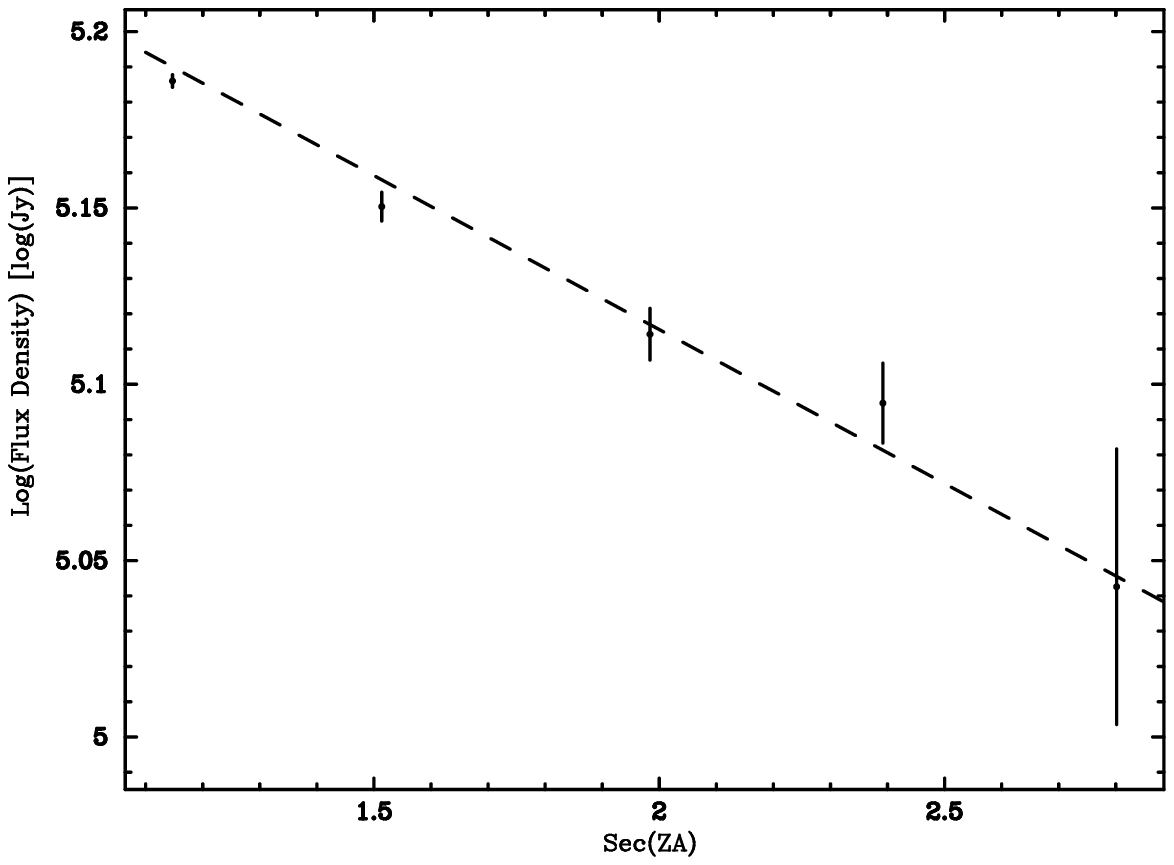}
\caption{\footnotesize 10 May 1998 - 16 May 1998 FLUX observations of Cas A, uncalibrated for
atmospheric attenuation.  The best straight-line fit of the natural log of the source flux density
to sec(ZA) is shown as a dashed line; this line corresponds to $\tau = 0.071$.  The
increased error bars at high airmass are due the smaller number of data per
bin, in addition to
the $\sim 1.6 \%$ gradient in the FLUX as a function of parallactic
angle discussed in the text.}
\label{fig:castau}
\end{figure}

Since the atmosphere has a significant opacity to microwave radiation, it is necessary
to apply a correction for atmospheric attenuation. The
observed flux density of a source as a function of zenith angle is
\begin{equation}
\label{eq:secza}
S\,(ZA) = S_o \, e^{-\tau \, {\rm sec}\, (ZA)};
\end{equation}
at 32 GHz, the optical depth $\tau$ is typically less than $0.1$ when observations are conducted.
Opacity corrections were determined by a straight-line fit of the natural logarithm
source flux densities to sec(ZA), and the data multiplied by $e^{+\tau \, {\rm sec}(ZA)}$.
Data at $ZA > 70^{\circ}$  were excluded from the analysis.  
 The 10 May 1998-16 May 1998 data set as a whole has
a mean optical depth $\tau = 0.071$.
Figure~\ref{fig:castau} shows the raw FLUX data
uncalibrated for atmospheric attenuation.  The straight-line fit in this figure corresponds
to $\tau = 0.071$.

\begin{figure}
\vspace{3.2in}
\includegraphics{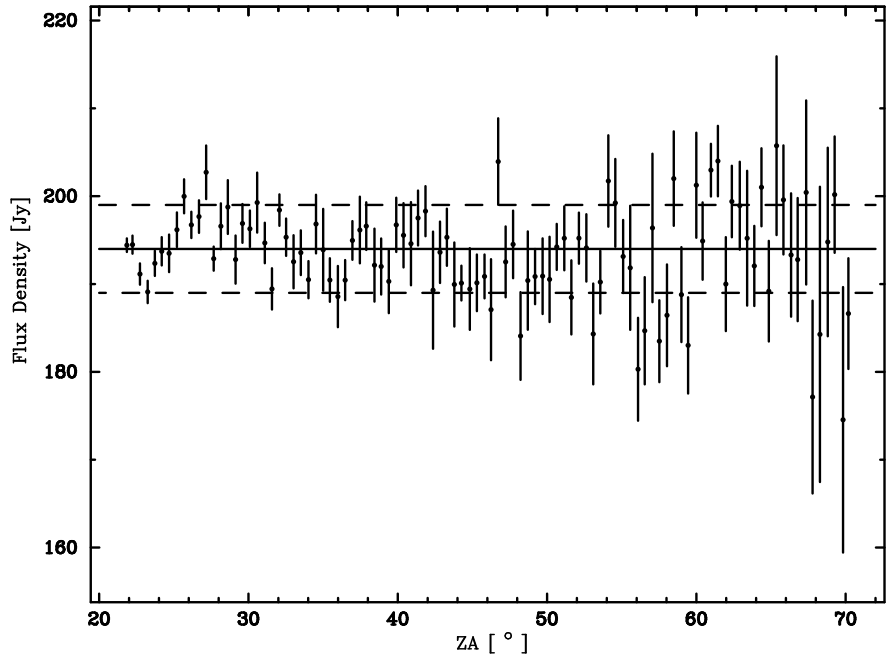}
\caption{10 May 1998 - 16 May 1998 FLUX observations of Cas A versus zenith
angle, calibrated for
atmospheric attenuation.  The feature at $  ZA \sim 28 $ is most 
likely  residual ground spillover ($\sim 1.5 {\rm \, mK}$).  
The solid line is the
average of all FLUXes; the dashed lines indicate the
range of values included at the $68\%$ confidence level using our estimated systematic
errors ($S_{\it cas} = 194 \pm 5$ Jy).}
\label{fig:casflux}
\end{figure}

\begin{figure}
\vspace{3.4in}
\includegraphics{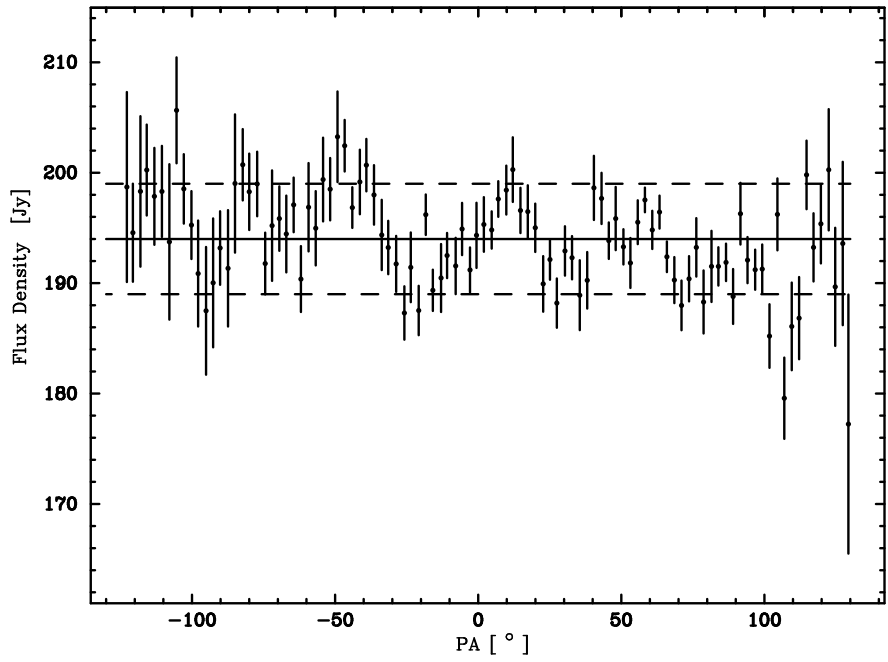}
\caption{10 May 1998 - 16 May 1998 FLUX observations of Cas A versus parallactic
angle, calibrated for
atmospheric attenuation.  
The slight overall gradient is due to an ambient temperature effect discussed in the text.}
\label{fig:caspaflux}
\end{figure}

The weighted mean of the calibrated data yields $S_{\it cas} = 194.0 \pm 0.4$
Jy (measurement error only). 
The fully calibrated data are plotted versus ZA in Figure~\ref{fig:casflux}.
The solid line shows the mean for all of the FLUXes, and the dashed lines show our
$1-\sigma\,$ error bars (see \S~\ref{sec:syserr}).  For clarity, the data have been binned
into 100 equal bins in ZA.
These data show evidence for some contamination at the $\sim 3$ Jy level; this
is likely due to ground spillover which was not entirely eliminated by our double-differencing.
An effect of similar magnitude is seen in the data plotted against parallactic angle (Fig.~\ref{fig:caspaflux}).  Exclusion of the data at $ZA > 50^{\circ}$ (data of notably lower quality) leaves the
mean unaffected to $<0.02\%$.
The overall gradient in the FLUXes as a function of parallactic angle is well-correlated with
ambient temperature and we therefore suspect that it is due to imperfect temperature
control of the cal diode; this was confirmed in calibration
observations taken in October and December of 1998.   Since the $T_{\it cal}$ measurements themselves
were taken at a wide range of ambient temperatures, this is not likely to
bias our result.  The fact that the the FLUXes as a function of ZA are roughly consistent with a constant
value ($\chi_{\nu}^2 = 1.62 \,$ for 89 degrees of freedom) is an indication that the parallactic
angle variations average out over the course of a track.
In any case both of these effects are well within our calculated
$1-\sigma\,$ systematic error limits, which we describe in the next section.

\section{Systematic Error Budget}
\label{sec:syserr}

In this section we present our estimate of the contributions of various systematic
effects to our measurement.  We consider these effects in order of their importance.

The dominant source of uncertainty for our measurement is the NIST gain
measurement error ($1.4\%\,$).  Short of recalibrating the telescope, there is
little that can be done to reduce this.  
Due to the differing reference planes of the NIST gain measurement and our hot/cold load
 calibration (see \S~\ref{sec:nist}),
there  is an additional  uncertainty which we estimate to be about 1\% due to ohmic
losses in the feeds.  This would have the effect of making the NIST gain 
artificially low,  hence affects our inferred flux density only in the negative 
direction.

Based on the rms of the fluxes derived from individual
days' observations we estimate that $\tau$ contributes
no more than a $1.0\%$ potential error to our measurement.  

Since the observed ground spillover and $T_{\it cal}$ effects noted in \S~\ref{sec:obs}
will average
out to some extent over a track (and as $T_{\it cal}$ measurements are
taken at different ambient temperatures), we take these to 
contribute independent systematic errors at the level of half the
 observed maximum
departure from the mean.  Since the observed fluctuations are $\sim 3$ Jy
in each case, this results in two contributions of $0.8\%$ to our systematic
error budget.

Since Cas A is located very close to the Galactic Plane ($ l = 111^{\circ}.74  , 
b = -2^{\circ}.128 $), there is the possibility that our measurements are
contaminated
by foreground sources, such as bright HII regions associated with young stellar systems,
as well as by flat spectrum background radio galaxies present at all galactic latitudes.
  Unfortunately Cas A is the
second brightest extrasolar source in the sky at our observing frequency, 
with the consequence that radio surveys
of this region of the sky tend to be heavily contaminated due to the telescope sidelobes; on
the other hand, the high overall flux density of Cas A reduces the fractional significance of interloping
sources.  The primary concern for observations with our dual-feed system is that a contaminating
point source may pass through the scan pattern of one of the reference beams.
  Since the reference
beams move on the sky relative to the nominal pointing center (Cas A), such an event introduces
a characteristic Hour Angle dependence into our data.  Based on the data
that we have collected, we place an upper limit of 16 Jy on the magnitude of possible point
source contamination in the reference beams.  We should indicate that while some of the
individual features in Figure~\ref{fig:caspaflux} are consistent with the signature of a single 
point source, all of them {\it jointly} are not.  In particular, the feature centered at
$PA = -20^{\circ}$ is consistent with a contaminating point source of about 16 Jy, but the 
double feature (including the feature centered at $PA = 33^{\circ}$) is {\it not} well represented
by two point sources.  We conclude therefore that most of the signal observed is residual ground
spillover.
Using a model which realistically accounts
for the motion of the reference beams during a scan, a 16 Jy source placed at the {\it center}
of one of our reference beams would bias the average flux density we infer for Cas A by $1.08$ Jy
($0.6 \%$).   We adopt this as a $1-\sigma$ upper limit on possible point source
confusion.  Since there are only a handful of sources in the whole sky at 32 GHz with flux densities greater than a few Jy,
this is likely to overestimate the level of point source contamination.

The accuracy of our pointing algorithm was determined using a Monte-Carlo
simulation with realistic signal-to-noise statistics.  The pointing error
is dominated by the model residuals, which could be improved by more
pointing in good weather.  The overall pointing error is estimated to
contribute a $0.7\%$ systematic error to our result, of which $0.5\%$ comes
from the pointing model residuals.

While determining the cal diode brightness temperature, $T_{\it cal}$, is one
of the most crucial aspects of our experiment, we have been able to do so
very precisely, so this error is one of the least significant contributors to our error
budget.
This is largely due to the fact that $T_{\it cal}$ is relatively insensitive
to errors in the load temperatures.  Using Equation~\ref{eq:tcalcold},
which takes into account the implicit load-temperature dependence of our system
temperature determination, we find that the error in $T_{\it cal}$ induced
by an error in the expected cold load temperature $T_{cold}$ is
\begin{equation}
\epsilon (T_{cal}) 
 = -\frac{y_{\it cal} - 1}{y-1} \times \epsilon (T_{cold})
\end{equation}
and the corresponding error due to the hot load is
\begin{equation}
\epsilon (T_{cal}) 
= \frac{y_{\it cal} - 1}{y-1} \times \epsilon (T_{hot})
\end{equation}
With the observed vales of $y \sim 2.71$ and
 $y_{\it cal} \sim 1.11$ , this yields about $0.6\%$ of error
in $T_{\it cal}$ per degree K of error in either of the hot or cold loads. 
 Adopting $\epsilon(T_{hot}) = 0.5\,{\rm K}\,$ and 
$\epsilon(T_{cold}) = 0.7\,{\rm K}\,$ and adding these errors in quadrature, we obtain a
$0.5\%$ error in $T_{\it cal}$.

Even though there is a fairly significant nonlinearity in our receiver system, it
affects our $T_{ant}$ calibration at only the $2\%$ level and so even relatively
larger errors in the characterization of the nonlinearity 
affect us at $\lesssim 0.4 \% \,$ level.  Systematic
errors in $T_{hot}$ or $T_{cold}$ do not significantly affect this characterization since 
the nonlinearity measurements depend only on the observed cal 
signal {\it increment} against each of these loads.

In general the gain and source spectrum may vary across the band, and this may
be a significant source of error for observations with a wide bandpass.  
For a source with a flux density $S(\nu)$ observed on a system with a frequency-dependent
 effective area $A_{eff}(\nu)$, the band-averaged spectral power is
\begin{equation}
\overline{P_{obs}} = \frac{1}{\Delta \nu} \int_{\nu_o -\Delta \nu/2}^{\nu_o +\Delta \nu/2} \,\,\, d\nu \, S(\nu) \, A_{eff}(\nu),
\end{equation}
where $\nu_o$ is the band center and $\Delta \nu$ the total bandpass.  If both $A_{eff}(\nu)$ and
$S(\nu)$ can be reasonably approximated as power laws over this interval, then the band-averaged
spectral power can be shown to be
\begin{equation}
\overline{P_{obs}} = A(\nu_o) \times S(\nu_o) + {\it O}\left(\frac{\Delta \nu}{\nu_o}\right)^2.
\end{equation}
Variations in these quantities thus induce errors 
proportional to the fractional bandpass
$\Delta \nu / \nu_o$ squared in the flux densities and brightness temperatures
one infers under the assumption that one is measuring $S_{\nu}(\nu_o)$ instead
of the band averaged flux density.  For our narrow ($2\%\,$) bandpass, this introduces
errors at less than the $0.1\%\,$ level and may be safely neglected.

Another potential concern for very accurate measurements made  with dual-feed systems
 is that the sidelobes of one beam will be present at a significant level in the
 other beam, reducing
the average gain of a doubly-switched measurement.  Since the secondary
is illuminated asymmetrically, the sidelobes of one beam in the direction
of the other tend to be highly suppressed.  The NIST measurements indicate that
the sidelobes of one beam are suppressed by $\gtrsim 35\,$ dB near the maximum
of the other beam (NIST, 1990); in contrast, the nearest sidelobes on the other side are at 
$-20 \,$ dB.  This is therefore not a concern for our system.

Our systematic error budget is summarized in Table~\ref{tbl:abssyserr}.
Adopting our measured value of $S_{\it cas} = 194.0 \pm 0.4$ Jy and the above
calculation of our systematic error  we
have $S_{\it cas,1998} = 194 \pm 5$ Jy at 32.0 GHz.

\begin{table}[h!]

\begin{center}
\begin{tabular}{ l l l r } 
 
\tableline\tableline
Source & Sense & Fractional Magnitude \\   \tableline
Nonlinearity & $\pm$ & $0.2 \%$ \\  
$T_{\it cal}$ (systematic)& $\pm$ & $0.5\%$ \\  
$T_{\it cal}$ (meas.)& $\pm$ & $0.5\%$ \\  
Point Source Confusion & $+$ & $0.6\%$ \\  
Pointing & $+$ & $0.7\%$ \\  
Unsubtracted Ground Spillover & $\pm$ & $0.8\%$ \\  
$T_{\it cal}$ Variations & $\pm $ & $0.8\%$ \\  
Atmospheric Opacity & $\pm$ & $1.0 \%$ \\  
Ohmic Losses in Feeds & $-$ & $1.0 \%$ \\
Antenna Calibration & $\pm$ & $1.4\%$ \\   
Total &  & $\pm 2.4\%$ \\  \tableline
\end{tabular}
\end{center}
\caption{Total Systematic Error Budget for 1.5-m Absolute Flux Measurement of Cas A}
\label{tbl:abssyserr}
\end{table}

\clearpage
\section{5.5-meter Observations and Leveraged Flux Density Scale}
\label{sec:extendedscale}

While Cas A is a useful calibrator for experiments with very large beams ($\theta_{\it FWHM} \gtrsim \theta_{\it cas A} \sim 5'$),
it is not so useful for instruments with smaller beams, or for experiments at southern latitudes.
It is therefore desirable to apply our absolute flux density measurement
 to a set of calibrators of smaller angular
extent at a wide range of declinations.
To this end, we have analyzed five epochs of observations of a set of calibrators (the choice of which is discussed below) relative
to contemporaneous observations of Cas A.  These observations were taken with the OVRO 5.5-meter telescope outfitted
with  a 32-\GHz Dicke-switching receiver system with a $\sim 6$ \GHz bandpass at 32 \GHz.  The primary beam of this instrument
has two beams with a FWHM of 7'.35 and a separation of 22'.17.  Using these observations plus our absolute value for $S_{cas}$
 we can establish an absolute flux density scale at our observing frequency for the set of sources we observed.

\begin{table}
\begin{center}
\begin{tabular}{ l l l }   
\tableline\tableline
Epoch  & Dates        & HEMT ID\\   \tableline
1      & 21dec94 -- 23dec94 & NRAO A-12 \\  
2      & 03feb95 -- 07feb95 & NRAO A-12 \\  
3      & 20apr95 -- 05may95 & NRAO A-12 \\  
4      & 29nov97 -- 09dec97 & NRAO A-23 \\  
5      & 22oct98 -- 28oct98 & MAO-5 \\   \tableline
\end{tabular}
\end{center}
\caption{Epoch definitions for source observations reported in this section, along
with the HEMT in place in our receiver during the observations.  See Table~4
 for the receiver band characteristics for each HEMT.}
\label{tbl:epochs}
\end{table}

The epochs of observation were chosen from archival OVRO data with the criteria that the source observations consist of 
observations of any of our sources close to observations of Cas A, that the observations occur over a reasonably wide range
of Hour Angles and that the observing conditions be sufficient
to obtain a reasonabe signal-to-noise on the source(s) over the observation as a whole.  The data we selected fall into
five epochs from December 1994 to October 1998;  these epochs, together with the serial number of the HEMT present in the
receiver at the time of the observations (see discussion below) are shown in Table~\ref{tbl:epochs}.

\begin{table}
\begin{center}
\begin{tabular}{ l l l }   
\tableline\tableline
HEMT ID   & $\nu_{1} \, ({\rm GHz})$ & $\Delta \nu \, ({\rm GHz})$ \\  \tableline
NRAO A-12 & 31.9 & 6.05  \\  
NRAO A-23 & 31.5 & 6.62  \\  
MAO-5     & $31.9 \pm 0.3$ & $6.6 \pm 0.6$  \\   \tableline
\end{tabular}
\end{center}
\caption{Bandpass characteristics of the 5.5-meter receiver with the three HEMTs used in observations.}
\label{tbl:hemt}
\end{table}

Our choice of sources was dictated by the requirement that the calibrators be bright, relatively
small in angular extent, and (for the most part) non-variable.  The sources we chose are NGC7027, 3C48, 3C84 ,
3C84 , 3C147, 3C218 , 3C273, 3C286, 3C345, 3C353, Jupiter,
Saturn and Mars.  The sources 3C48, 3C147 and 3C286 are quasars with steep spectra, and are regularly observed
at the VLA as primary flux calibrators.  The latter sources, and in particular Jupiter and Mars, are frequently
 used calibrators at centimeter wavelengths.   The possible variability of some of these sources will be discussed below.

Since Cas A is comparable in size to the 5.5-meter primary beam, it is necessary to correct the observed
flux densities for the effect of the convolution of the sky brightness with the telescope beam pattern.  The details of
this procedure are discussed by Leitch et al. (1999) and Leitch (1998). In brief, a high-resolution map of Cas A made at
32 GHz on the Effelsberg telescope (Morsi 1997) was used as a brightness template in a Monte Carlo simulation,
incorporating uncertainties in the 5.5-meter beam pattern and pointing model, to determine the ratio $f$ of
the total flux density of Cas A to the flux density observed with the 5.5-meter on Cas A .  The $68\%$ confidence
interval of the resulting distribution is $f=1.18^{+0.02}_{-0.01}$.  This is the factor by which the 5.5-meter
observed $S_{\it cas}$ values must be multiplied to obtain the total flux density of Cas A, as would be observed by a
telescope with a primary beam much larger than the extent of Cas A itself.  This allows us to relate the
5.5-meter observations of alternate calibrator sources {\it relative} to Cas A to our {\it absolute} measurement
of Cas A with the 1.5-meter telescope.

Given the wide bandpass of the 5.5-meter receiver system the
variation of the source spectra across the observed band and the bandpass of the observing band itself
must be accounted for.
Leitch (1998) has extensively characterized the band properties of the 5.5-meter receiver from 1994 - 1997.
Over this time period, two different HEMT's were employed in the receiver, and the receiver in each
of these states has been fully characterized separately.  In May of 1998, A-23 suffered a catastrophic
failure and was replaced with the MAP non-flight HEMT MAO5.  A full characterization of the receiver
with MAO5 in place has not been undertaken, but the data we have in hand are in
reasonable agreement with the bandpass of A23.  From these, we estimate
$\delta \nu = (6.6 \pm 0.6)$ \GHz , $\nu_{cent} = (31.9 \pm 0.3)$ \GHz for the receiver with MAO5.  
Our adopted band characterizations are summarized in Table~\ref{tbl:hemt}.

The strategy we choose is to interpolate the broadband 5.5-meter measurements of our calibrator sources
(relative to Cas A) onto a 32.0 GHz value using our narrow-band measurement of $S_{\it cas}$ at this frequency and
spectral information culled from the literature, or, for the case of Cas A, obtained from measurements at
OVRO. For 3C48, 3C147 and 3C286 we compute
effective spectral indices over the 5.5-meter bandpass from the spectral fitting formulae given by
(Kellermann et al. 1969), with a standard power law of the form
\begin{equation}
S(\nu) = S(\nu_o) \times \left(\frac{\nu}{\nu_o} \right)^{\alpha}.
\end{equation}
Our effective spectral indices for these sources are shown in Table~\ref{tbl:specind}.  Although the
formal errors in these spectral indices given the uncertainties quoted in Kellermann's fitting formulae
are of order 0.01, we adopt error bars of 0.1 in order to allow for the extrapolation to
32 \GHz and the possibility of secular evolution of the spectra at high frequency.
For NGC 7027, a planetary nebula, we assume $\alpha = -0.1 \pm 0.1$; and
for the other 3C sources , we assume that $\alpha = -0.8 \pm 0.1$.  
Kellermann
finds from low frequency ($\nu < 15$ \GHz) observations of Cas A that the spectrum
is well described by a single power law with $\alpha_{cas} = -0.765 \pm 0.005$.  Rather than extrapolate
this low frequency result to 32 \GHz, we use the 14.5 and 31.7 \GHz RING5M measurements of
 Cas A (Leitch et al. 1999) to compute
a spectral index which is more representative of the actual spectrum at our frequency.  The RING5M measurements
result in $S_{\it cas,14.5} = 313.6 \pm 8.9 {\rm \, Jy}$ and $S_{\it cas,31.7} = 164.2 \pm 5.4 {\rm \, Jy}$, implying
$\alpha_{cas} = -0.827 \pm 0.056$.  This result is slightly steeper than the result
of Kellermann at lower frequencies, but still consistent with a single power law over the entire range up
to 31.7 \GHz. These measurements are also in good agreement with the results of recent observations of
Cas A with the BIMA array (Wright et al. 1999).  In this analysis, Wright et al. have combined BIMA data at 28 and 83 \GHz with
1 and 5 \GHz VLA maps of Cas A (Koralesky and Rudnick, 1999) to examine spectral index variations of the various ``knots'' of emission
seen in Cas A at high resolution maps.  The spectra of these structures are generally well described by 
power laws with $-0.81 \lesssim \alpha \lesssim -0.77$, up to and including the 83 \GHz data.  Since
a single dish radio telescope is likely to measure slightly different spectral indices due to the dependence
of the spectral indices on spatial scale, we take this as merely suggestive, and
adopt the value of $\alpha_{cas}$ computed above to interpolate our 5.5-meter measurements
onto 32.0 GHz flux densities for all of our sources.  We approximate the spectrum of Jupiter across
our band using the model of Trafton (1965) as presented in the context of Wrixon et al.'s (1971) measurements.
Using this model, and assuming a constant power-law slope across our band, we compute an effective
spectral index of $2.24$ from 28.5 to 40.0 GHz, which we use in relating our band-averaged fluxes
to 32.0 GHz flux densities; we estimate a 1-$\sigma$ error of 0.1 in this quantity.
For the other planets we assume a thermal ($\alpha = 2$) spectrum, with the same uncertainty.
This strategy should do a good job of accounting for the effects of the overall shape of the
spectrum on the average brightness temperature from 29 to 35 GHz, but may not get the actual 32.0 GHz
brightness temperature for sources with very narrow features in the spectrum (like Mars).

\begin{table}[t!]
\footnotesize 
\tablewidth{6.5in}
\centering
\begin{tabular}{ l l }   
\tableline\tableline
Source & Spectral Index \\   \tableline
Cas A & $-0.827 \pm 0.056$ \\  
NGC 7027 & $-0.1 \pm 0.1$ \\
3C48 & $-1.18 \pm 0.1$ \\  
3C147 & $-1.21 \pm 0.1$ \\  
3C286 & $-0.82 \pm 0.1$ \\  
Jupiter & $+2.24 \pm 0.1$ \\  
Saturn & $+2.0 \pm 0.1$ \\  
Mars & $+2.0 \pm 0.1$ \\   \tableline
\end{tabular}
\caption{Spectral Indices used for the interpolation from wideband flux measurements
to 32.0 GHz flux densities.   See the text for details on how the indices and uncertainties
were estimated.}
\label{tbl:specind}
\end{table}

It is well known that the flux density of Cas A is slowly decreasing as a function of time.
Baars (1977) gives an expression for the fractional annual decrease in Cas A's flux density:
\begin{equation}
\frac{\delta S}{S} = [ (0.0097 \pm 0.0004) - (0.0030 \pm 0.0004) \, {\rm log} \, \nu_{\rm GHz} ] \, {\rm yr^{-1}} ;
\end{equation}
where $\Delta t$ is the time interval in years.  $S_{\it cas,t}$ is then
\begin{equation}
S_{\it cas,t} = S_{\it cas,t_o} \times (1 - \frac{\delta S}{S})^{\Delta t}.
\label{eq:tcas}
\end{equation}
At $\nu = 32\,{\rm GHz}$, this yields a $(0.52 \pm  0.07) \%\,$ decrease in $S_{\it cas}$ per year.
This functional form for the secular variation has found some support at 15 GHz
in the work of O'Sullivan and Green (1999).
We take $\frac{\delta S}{S} = 0.52 \pm  0.15 \,  {\rm yr^{-1}}$. 
Using this expression we can compare observations taken at
differing epochs to our absolute measurement of Cas A  (epoch May 1998).
For the epochs we are examining this
is at most a 2\% correction. A misestimation of the variability of Cas A will increase the variance
of  the flux ratios we observe for a given source, and hence  be automatically accounted for in the final error budget
of each source.

For these observations, scans on these calibrators were interleaved with observations of Cas A over
a wide range of Hour Angles, allowing an accurate determination of the atmospheric opacity and frequent relative calibration using Cas A.  
Table~\ref{tbl:relflux} shows the average observed flux ratios of our sources for all epochs.  Where the epoch-to-epoch 
scatter in the observed ratios exceeds the mean error bar for a given source, the epoch-to-epoch scatter
is adopted as the error in the ratio.  This strategy automatically incorporates into our final error bar any systematic
errors which may not be manifest in the errors estimated for a given epoch's observations, such as source
variability and any errors in the characterization of Cas A's secular behavior.
For each source ``X'' with spectral index 
$\alpha_x$, measured relative to a calibrator ``C'' (with associated flux density $S_{c,32}$ and spectral
index $\alpha_c$) with a rectangular bandpass $\delta \nu$ centered at a frequency $\nu_1$,
the narrow-band flux ratio referenced to an epoch $t_o$, $R_{32.0,t_o}$, can be shown to be
\begin{equation}
R_{x,32.0,t_o} =  R_{x,obs,t} \, \times \, (1 - \frac{\delta S}{S})^{(-\Delta t)} \times \frac{\alpha_x+1}{\alpha_c+1} \, \times
\frac{ \nu \left(\frac{\nu}{32} \right)^{\alpha_c} \mid^{\nu_1+\delta \nu/2}_{\nu_1-\delta \nu/2}
}{  \nu \left(\frac{\nu}{32} \right)^{\alpha_x} \mid^{\nu_1+\delta \nu/2}_{\nu_1-\delta \nu/2} },
\label{eq:rr}
\end{equation}
where the values of $\Delta t$ , $\nu_1$  and $\delta \nu$ used are those appropriate to the epoch of observation.
These are the ratios we quote in Table~\ref{tbl:relflux}.
While the data have been corrected for atmospheric attenuation
using a mean optical depth $\tau = 0.049$, these corrections are relatively unimportant for our measurements
since any errors in $\tau$ will tend to cancel in the ratio.   
The flux ratios quoted have also been corrected for
the form factor indicated above, and the error bars include a $1.7\%$ systematic error due to the determination
of this form factor.  Planetary fluxes have been expressed in K/Jy using the relation
\begin{equation}
\frac{T}{S_{\it cas}} = \frac{\lambda^2 R}{2 k \, \delta \Omega} , 
\end{equation}
where $T$ is the brightness temperature of the planet, $\delta \Omega$ is the solid angle
of the planet at the time of the observation (determined from standard ephemerides), $R$ is
the observed ratio of the planetary flux to the flux of Cas A, and $k$ is Boltzmann's constant.
For this calculation, we used the value of $\lambda$ appropriate to each epoch of observation.

With these flux ratios in hand, it is then straightforward to obtain 
the flux density at 32.0 \GHz $S_{x,32}$ of the source ``X'' :
\begin{equation}
S_{x,32} =  R_{x,32,t_o} \, \times \, S_{c,32,t_o} .
\label{eq:sx}
\end{equation}
For the case at hand, $S_{c,32,t_o} = 194 \pm 5$ Jy.  These results are also shown in Table~\ref{tbl:relflux}.

\begin{table}
\footnotesize 
\tablewidth{6.5in}
\centering
\begin{tabular}{ l l l l }   
\tableline\tableline
Source & Epochs & Mean Flux Ratio & $S_{\nu}(32)$ or $T(32)$ \\  \tableline
NGC7027 & 1-3 & $(2.76 \pm 0.02) \times 10^{-2}$ & $ 5.50 \pm 0.20$ Jy \\  
3C48   & 2,5    & $(4.37\pm 0.40)\times 10^{-3}$ &  $ 0.86 \pm 0.08 $ Jy\\  
3C84   & 4,5    & $(6.22\pm 0.03)\times 10^{-2}$ &  $ 12.3 \pm 0.4 $ Jy\\  
3C147  & 1-5    & $(7.29 \pm 0.28) \times 10^{-3}$ &  $ 1.44 \pm 0.08 $ Jy\\  
3C218  & 2,3    & $(9.55 \pm 0.28) \times 10^{-3}$ &  $ 1.88 \pm 0.06 $ Jy\\  
3C273  & 4,5    & $(2.14 \pm 0.02) \times 10^{-2}$ &  $ 42.4 \pm 1.6 $ Jy\\  
3C286  & 2-5    & $(9.83 \pm 0.42) \times 10^{-3}$ &  $ 1.95 \pm 0.11 $ Jy\\  
3C345  &  5     & $(5.10 \pm 0.11) \times 10^{-2}$ &  $ 10.11 \pm 0.43 $ Jy\\  
3C353  &  2,3   & $(1.92 \pm 0.02) \times 10^{-2}$ &  $ 3.80 \pm 0.43 $ Jy\\  
Jupiter & 2-5 & $ 0.800 \pm 0.030 $ &  $ 155.1^{+6.8}_{-5.7} $ K\\  
Mars   & 2,3    & $ 0.990 \pm 0.028 $ &  $ 196.0^{+7.5}_{-7.6} $ K\\  
Saturn & 4,5    & $ 0.726 \pm 0.019 $ &  $ 140.8^{+4.4}_{-4.9} $ K\\    \tableline
\end{tabular}
\caption{Observed ratios of calibrator sources relative to Cas A.  Planets' relative fluxes
are expressed in units of K/Jy.  All fluxes have been corrected to epoch May 1998 on an
epoch-by-epoch basis, and interpolated onto narrowband flux ratios using Equation~19.}
\label{tbl:relflux}
\end{table}

Since in general a Gaussian distribution of spectral indices does not give rise to a Gaussian
distribution of flux densities, Equations~\ref{eq:rr} and ~\ref{eq:sx} were jointly evaluated $10^6$ times for each source ``X'' with
independently varying $R$, $\alpha_c$ , $\alpha_x$ , $S_{c,32}$ , $\nu_1$  and $\delta \nu$ values drawn
from the (assumed Gaussian) populations implied by the error estimates we have quoted above.  For $R$, we
have used {\it each epoch's} observed flux ratio together with the error bar inferred from the scatter internal to those
data.
For sources not bright enough to point on, we also included
the effects of the residuals in the 5.5-meter pointing model (0'.45 rms -- an average over all epochs).
The error bars quoted in Table~\ref{tbl:relflux} are the 68\% confidence intervals inferred from this simulation.
For the 3C sources and NGC7027, the error intervals
inferred are virtually unchanged relative to those which are obtained by ignoring finite
bandpass effects.  This is due to the fact that these sources have
spectral indices very similar to that of Cas A.  

\begin{figure}
\vspace{3.4in}
\includegraphics{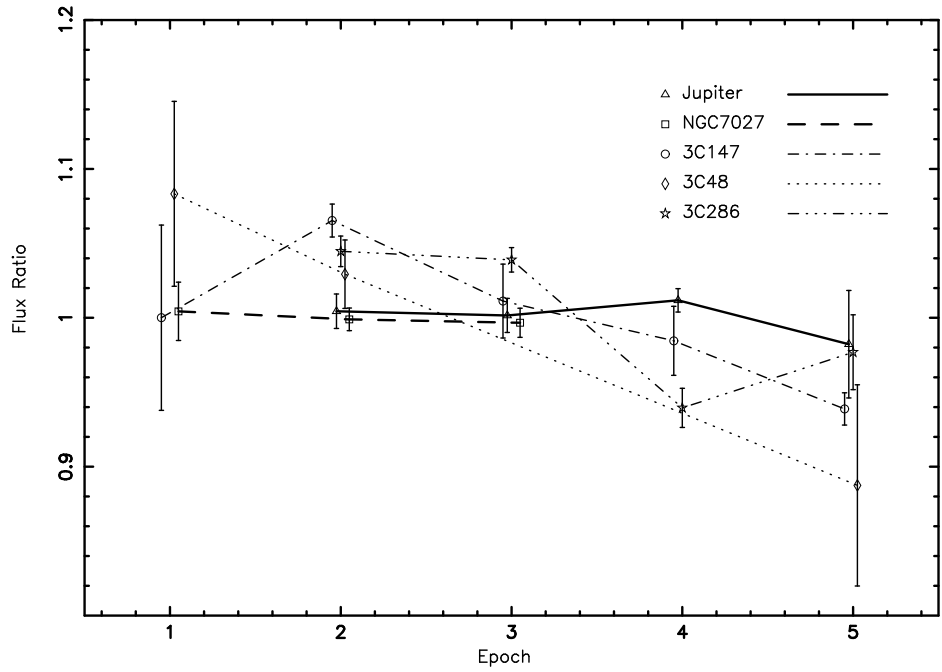}
\caption{\footnotesize Flux ratios for the best potential calibrator sources in our sample relative to
Cas A.  The flux ratio for each epoch of each source has been divided by the mean flux ratio for that source;
Jupiter has been corrected for the bandpass characteristics quoted in Table~\ref{tbl:hemt}.  All sources
been corrected for the secular evolution of Cas A.}
\label{fig:srcratios}
\end{figure}

Figure~\ref{fig:srcratios} shows the {\it relative} flux ratio for the five most-observed sources in our sample as a function of
epoch.  While there is some indication that epochs 4 and 5 may be slightly lower than other epochs, it is difficult to 
disentangle this effect from the intrinsically large scatter of the data.  In particular Jupiter, the brightest of these 
five sources,
shows no departure from the mean greater than 2\% in four epochs of observation; NGC7027, the second brightest, shows
less than this.  The data points for these two brightest sources are shown connected by bold lines in the figure; they are
consistent with a constant value of 1.0.  For Jupiter, the uncertainties in the source spectra and band characterization
have been included in each epoch's error bar;  these uncertainties dominate the uncertainty in the epoch 5 flux ratio.

Some of these sources are at least slightly variable,
and the variability of some of them is not known.  Based on our observations and existing data, we designate the planets,
plus NGC7027, 3C48, 3C147, and 3C286 as useful calibrators at 32.0 GHz.  There are indications that
3C147 may be slightly variable.
For convenience, we summarize our findings for these sources in Table~\ref{tbl:calib}.  Our brightness temperature for
Mars has been corrected from a mean heliocentric radius of 1.658 AU for the epochs of observation to a fiducial heliocentric
radius of 1.524 AU using an $r^{0.25}$ law (Epstein, 1971) to facilitate comparison 
with other measurements.  In this table we have adopted the more precise Leitch (1998) 
values for $S_{3c286}/S_{cas}$ and $T_{Jup}/S_{cas}$.  

\begin{table}
\begin{center}
\begin{tabular}{l l}
\hline\hline
Source & $S_{\nu}\,$ (Jy) or $T$ (K) \\\hline
Cas A$\phantom.^{\rm A}$ & $(194 \pm 5 {\,\rm Jy})$ \\  
NGC7027 & $5.45 \pm 0.20 {\,\rm Jy}$ \\  
3C48 & $0.86 \pm 0.08 \, {\rm Jy}$ \\  
3C147 & $1.44 \pm 0.08\, {\rm Jy}$ \\  
3C286 & $2.02^{+0.05}_{-0.06} \, {\rm Jy}$ \\  
Jupiter & $152 \pm 5 \, \, {\rm K}$ \\  
Saturn & $141^{+4}_{-5} \, \, {\rm K}$ \\  
Mars$\phantom.^{\rm B}$ & $200 \pm 8 \, \, {\rm K}$ \\
\hline
\multicolumn{2}{l}{$\phantom.^{\rm A} \,$ 32.0 GHz flux density standard.} \\
\multicolumn{2}{l}{$\phantom.^{\rm B} \,$ $T_{Mars}$ corrected to 1.524} \\
\multicolumn{2}{l}{\phantom{BB} AU heliocentric radius;}\\
\multicolumn{2}{l}{\phantom{BB} see text for details.} \\
\end{tabular}
\end{center}
\caption{32.0 GHz Calibrator Flux Densities and Brightness Temperatures \label{tbl:calib}}
\end{table}

\clearpage

\section{Discussion and Conclusion}
\label{sec:concl}

We have presented an accurate measurement of the 32 GHz flux density of Cas A, $S_{cas,1998} = 194 \pm 5 \,\,{\rm Jy}$ .
This is the first direct measurement of $S_{\it cas}$ above 16 GHz.
This measurement provides
a firm foundation for calibrating experiments in the vicinity of 32 GHz.  The resulting flux
scale has an uncertainty of $2.5 \%$, a significant improvement over other flux density scales at 
this frequency, most of which contain uncertainties $ > 6\%$.

By way of comparison, the RING5M program conducted at OVRO recently
derived $S_{\it cas}$ relative to observations of DR21 on the OVRO
5.5-m telescope at 32 GHz (Leitch et al., 1999).  Using Dent's (1972)
flux density for DR21, and applying the form factor for the OVRO 5.5-m
beam derived from a 32 GHz map of Cas A made with the Effelsburg
telescope (see \S~\ref{sec:extendedscale}) , they find $S_{\it
cas,1996} = 195.8^{+6.4}_{-5.7}$ Jy, corresponding to $S_{\it
cas,1998} = 193.8^{+6.4}_{-5.7}$ Jy.  This is in agreement with the
result reported in this paper.

Until the RING5M calibration work, the OVRO Ka-band flux density scale
was based on the Wrixon et al.(1971) absolute measurement of Jupiter,
$T_J = 144 \pm 8\,$ K.  
This implies  $S_{\it cas,1996} = 185.5 \pm 10.3\,$ Jy, consistent with both this
absolute measurement and the analysis relative to DR21 (although the latter is not completely
independent of it), but with an associated error of $5.5\%$, more than twice that
in our measurement of Cas A.  The Kellermann scale implies $S_{\it cas,1994} = 186.5 \pm 3.7 $, including
only formal errors in the Kellermann scale, but with a much larger error associated with
the extrapolation to 32 GHz and the secular correction.  The Saskatoon experiment
(Netterfield, 1997) estimated an uncertainty of $13\%\,$ in this value. 

We have also presented absolute flux densities for a standard set of
centimeter-wavelength calibrator sources, including Jupiter and Mars.
These measurements should facilitate the calibration of experiments at
a wide range of angular scales and terrestrial latitudes.  These
measurements have $1-\sigma$ accuracies ranging from $3\%$ (for
Jupiter) to $9\%$ (for 3C48).  Our measurement for 3C286 is also in
good agreement with that of Leitch (1998), who finds $S_{3C286}(32
\GHz) = 2.02 \pm 0.07$.  Our inferred brightness temperature for
Jupiter ($T_J = 152 \pm 5 \,$ K) is in excellent agreement
with both Wrixon's $T_J = 144 \pm 8 \,$ K, and the RING5M (Leitch et al.,
1999) value of $152 \pm 5 \,$ K.  We emphasize that the Leitch 
value of $T_J$ is independent of our quoted value insofar as the
absolute calibrations of these experiments were independent.  Our measurement
of $T_{\it Mars} = 200 \pm 8\, {\rm K}\,$ is consistent
with the results of Hobbs and Knapp (1971), who find $T_{\it Mars} =
207 \pm 13 {\rm \,\, K}\,$ at 9.55 mm, and with Ulich (1981) who finds
$T_{\it Mars} = 194 \pm 8.2 {\rm \,\, K}\,$ at 9.5 mm.  All of these
results are referred to a standard solar distance of 1.524 AU. The
compilations of microwave observations of Mars given by Epstein (1971)
and Efanov et al. (1971) are also in good agreement with our
measurement near 1 cm; our results are significantly more precise than
any of the 1-cm measurements reported in these compilations.  Our
measurements of these sources offer an excellent basis for the
calibration of current and future experiments in the vicinity of 32
\GHz.

This work would not have been possible without the expertise of the
OVRO staff; in particular, we acknowledge the assistance of Russ
Keeney for his work on both the 1.5-meter and 5.5-meter telescopes,
and Stan Hudson for constructing the cold load box.  
We thank the OVRO Millimeter Array staff for
useful discussions and periodically loaning us equipment, and
Ken Kellermann, Tim Pearson, and Ed Wollack for comments on the manuscript which 
improved the content and presentation.  We acknowledge
NRAO, Princeton University, and the MAP collaboration for loaning us
the HEMT used in the October 1998 5.5-meter observations.  For part of
the duration of this work, BSM was supported by the Zacheus Daniels
fund at the University of Pennsylvania.  STM was supported by an 
Alfred R. Sloan Fellowship at the University of Pennsylvania.  The work was 
supported in part by NSF grant AST-9419279.

\newpage

\end{document}